# Creating Geospatial Trajectories from Human Trafficking Text Corpora

Saydeh N. Karabatis and Vandana P. Janeja

Department of Information Systems, University of Maryland Baltimore County

{saydeh1, vjaneja}@umbc.edu

Human trafficking is a crime that affects the lives of millions of people across the globe. Traffickers exploit the victims through forced labor, involuntary sex, or organ harvesting. Migrant smuggling could also be seen as a form of human trafficking when the migrant fails to pay the smuggler and is forced into coerced activities. Several news agencies and anti-trafficking organizations have reported trafficking survivor stories that include the names of locations visited along the trafficking route. Identifying such routes can provide knowledge that is essential to preventing such heinous crimes. In this paper we propose a Narrative to Trajectory (N2T) information extraction system that analyzes reported narratives, extracts relevant information through the use of Natural Language Processing (NLP) techniques, and applies geospatial augmentation in order to automatically plot trajectories of human trafficking routes. We evaluate N2T on human trafficking text corpora and demonstrate that our approach of utilizing data preprocessing and augmenting database techniques with NLP libraries outperforms existing geolocation detection methods.

## 1 INTRODUCTION

Human Trafficking is an illegal business that includes recruiting, transporting, harboring people through deceit, coercion, or fraud with the objective of exploiting them to make profit. It is a modern-type of slavery and epidemic crime that affects the lives of millions of people across the globe [1]. Various domestic and international laws exist that prohibit trafficking and protect the victims [2]. In the US, the Trafficking Victims Protection Reauthorization Acts (TVPRA) established methods to persecute traffickers, protect victims of trafficking, and "respond to disaster areas where people are particularly susceptible to being trafficked" [3]. Members of law enforcement agencies have been working tirelessly to disrupt such operations and free the enslaved victims. Anti-trafficking humanitarian organizations offer services to survivors to help them heal. Many narratives posted on media and anti-trafficking organizations sites contain a varying number of details about the experience of survivors during their captivity. Those narratives are either written by journalists or by members of the humanitarian organizations following interviews of the survivors. Based on the posted information, relevant agencies can hamper such activities, help shape policies to prevent these heinous crimes, and bring justice to the victims.

In order to protect the freed victims, most media sites often provide vague location names of the route travelled by the victims [4]. Very few narratives tell it all: survivor names, place of origin, visited places along the transportation routes, and the final destination. We were able to locate articles that narrate stories of victims and list the location names along the transportation route in chronological order. However, these articles do not disclose any personal information of the people involved in the trafficking event. We use these articles to show the efficacy and validity of our proposed system.

*Motivating Example: A group of Syrian migrants traveling described their smuggling journey from Syria to Lebanon, then riding the Mediterranean waves to a Syrian refugee camp in Turkey in 2017. Each migrant paid $2,500 to the smugglers to illegally cross the water border to Greece. They were jam-packed in a small inflatable boat that set sail towards Greece. Several migrants were scared once they reached the open water and tried to steer it back to Turkey, but the smugglers fired shots at them and forced them to stay the course. Several fell off the boat and drowned. The survivors navigated through the waves at night using their smartphone GPS app. Once they entered the Greek territorial waters, they punctured their boat and issued signals of distress. The Greek Coast Guard was able to rescue the remaining ones and brought them to the closest Greek island. These migrants witnessed abuse at the hands of their smugglers and paid outrageous amounts of money to illegally cross as short as a single mile of water.*

This is just one of the many examples of narratives documenting the journey of trafficking victims. One way law enforcement agencies can disrupt human trafficking activities is to predict the route of these movements by gaining knowledge about trafficking routes that took place. Non-profit organizations can reach out to human trafficking victims during their trafficking routes. Given the overwhelming number of survivor narratives reported on the internet and since very few of these narratives contain particular names of places of the transporting routes, it is not feasible for a human to read each narrative, extract and identify the location names, and plot on a map the travelled trajectory. Therefore, it is very important for law enforcement and anti-trafficking organizations to obtain a tool that will automatically read the narratives, extract the location names if available, and plot on a map the transporting route using the location names. Such tool will help the officials identify past trajectories and gain insight of future routes in order to combat these inhumane operations.

In this paper, we present a *Narrative to Trajectory (N2T)* prototype system that automatically reads the human trafficking narratives and traces the trajectory of their transportation route on a map. The contributions of the paper are as follows:

**1.** We present an AI model that identifies location names and their corresponding geolocations in order to automatically plot the trafficking trajectory as narrated in the text corpus on a map.

**2.** We compare the performance of *N2T* to existing Integer Linear Programming (*ILP*) and *NLP* tools (*SpaCy* and *NLTK* libraries, and *BERT*, a Deep Learning model). We demonstrate that our method performs better than *ILP, SpaCy,* and *NLTK*, and same as *BERT*.

The remainder of the paper is organized as follows: In Section 2, we present related work. In Section 3, we describe our *N2T* methodology. In Section 4 we present its experimental results and its performance evaluation measures. We compare *N2T* to existing *ILP, SpaCy, NLTK,* and *BERT* techniques. Section 5 provides conclusion and future work.

## 2 RELATED WORK

Some organizations that deal with human trafficking generate statistical analysis reports to bring awareness to the public, advice government agencies, and draft legislative policies [4, 5, 6, 7]; others offer counseling, shelter, and monetary funds to help the victims [8]. A lot of work has been done analyzing sex trafficking ads: Extracting text using Natural Language Processing (NLP) techniques from ads to identify telephone numbers of sex laborers [9], or to identify sex trafficking ads using a classifier [10], or use a deep learning model to analyze sex ads that contain both images and text [11], or extract clusters of data with prior history of human trafficking activities by scanning sex ads from the internet [12]. None of these works addresses mining geolocations from media articles.

Extracting geographical tags from websites containing illicit advertisements written in "unusual language models" is presented in [13]. The authors describe a framework that parses human trafficking advertisements and produces a set of geolocation names by using a context-based classifier and Integer Linear Programming (*ILP*) model in conjunction with the openly available GeoNames knowledge base. Although this approach is somehow similar to ours in leveraging GeoNames to identify location names, we differ in the steps of gathering, pre-processing and tokenization. We propose a system which shows improvement over *ILP* because of narrative pre-processing, the use of sentence tokenization, multi-word lexicon, and database techniques to identify geospatial tokens. Moreover, we trace the trafficking trajectory based on the reported narratives. To the best of our knowledge, this paper presents the first AI model for an automatic extraction of geospatial trajectories based on human trafficking narratives included in text corpora.

## 3 METHODOLOGY

### 3.1 Narrative to Trajectory

The goal of this research is to create a system that automatically extracts location names from a text corpus and uses geolocation coordinates to create geospatial trajectory as it was narrated. This system, called *Narrative to Trajectory* (*N2T*), accepts a narrative as input, preprocesses it, splits it into tokens, and if needed, labels each token according to its semantics and its syntax in the sentence. A *token*, the most atomic part of *N2T*, is a sequence of contiguous characters semantically grouped together. A token can be either a single word *(SW)* or a multi-word *(MW)* and is defined as:

*Definition 1:* A *token* $to_i^P$ is an element in a text corpus *To*, where $To = \{to_1^P ... to_i^P \mid to_i^P \in SW \vee MW\}$, where $SW = \{sw_1, ..., sw_i \mid |sw_i| = 1\}$, where $MW = \{mw_1, ..., mw_i \mid |mw_i| > 1 \text{ and } mw_i \subset SW\}$, where $P=(T, V, A, G, La, Lo)$, such that:

- A temporal index $T = \{1,2,....n\}$ where *n* is an index signifying the position of the token in the corpus
- A value $V = SW \vee MW$
- A tag $A = \{noun, verb, adjective, adverb ...\}$
- A geospatial flag $G = \{0, 1\}$ such that if $G = 1$ then token is geospatial, otherwise not.
- A latitude value *La* and a value longitude *Lo* when $G = 1$.

*Definition 2:* A *sentence* $SNT_i$ is an element in a set of sentences *SNT*, where $SNT = \{SNT_1... SNT_n\}$ where $SW \subset SNT$ and $MW \subset SNT$

*Definition 3:* A trajectory is defined as: $T_r = (g_1, g_2, ..., g_j)$, where $g_j$ is a sequence of geospatial token on the route and $g_j \in To$ such that $to_j^G = 1$. The terms trajectory and path are used interchangeably throughout this work.

A high level architecture of *N2T* is shown in Fig. 1. First, the narrative is read, preprocessed and split into tokens using a Multi-word lexicon (MW Lexicon), while preserving the order of token appearance in the narrative. The tokens are tagged according to their part of speech. *N2T* includes a geolocation database (Location Dimension) containing location names with their longitude and latitude as obtained from GeoNames [14]. Using Location Dimension, geospatial tokens are identified, and plotted on a map sequentially revealing the trajectory as described in the narrative.

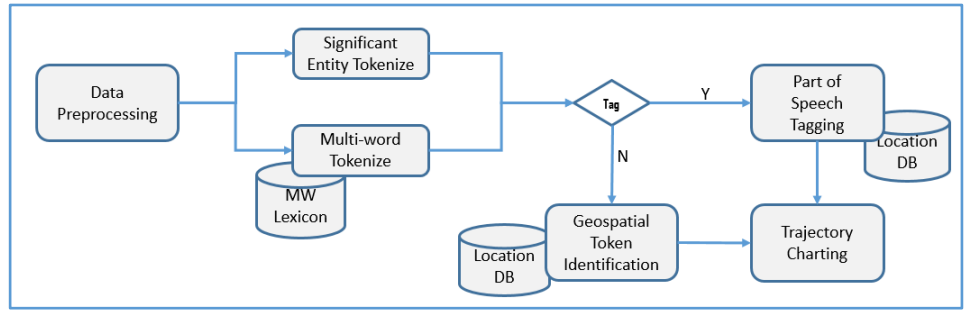

**Fig. 1.** Overview of the *N2T* System Architecture

**3.2 *N2T* System Flow**

In this subsection, we describe in detail the steps used to implement *N2T* system as presented in Fig. 1. In order to extract tokens, we applied two NLP libraries to the text corpus. Both libraries generated tokens and labeled them using grammar rules. However, both libraries failed to extract a large number of geospatial locations due to the presence of special and non-ASCII characters in several location names, multiple word location names, and different spellings of the same location name. For these reasons, data preprocessing must be performed before tokenization.

*3.2.1 Data Preprocessing*

We conduct the following data preprocessing steps to every narrative in the text corpora:
1. We identify special characters and replace them with characters that are NLP digestible.
2. We identify the non-ASCII characters and replace them with their equivalent ASCII characters.
3. We augment the MW Lexicon with new multi-word location names when found in the narratives.
4. We include all homonyms of the same location in the Location Dimension.

*3.2.2 Tokenization*

Tokenization splits the narrative into meaningful tokens. We use two techniques:
- Significant Entity Recognition Tokenization (ST): uses Spacy, it segments text into "potentially interesting tokens" by applying lexical semantic rules specific to the language of the text [15].
- Multi-word Tokenization (MWT): uses NLTK, it applies sentence tokenization to the input, then tokenizes each sentence into single tokens. Using the MW lexicon, MWT identifies tokens composed of multiple words [16].

*3.2.3 Geospatial Token Identification*

In this step, we describe the process of identifying the tokens that are locations. Using geospatial augmentation and database techniques, we augment the token list generated during tokenization with the Location Dimension and generate a list of geospatial tokens. The result is an ordered list [($to_i^V$, $to_i^{Long}$, $to_i^{Lat}$, $to_i^G$)], where $to_i^V$ is the geolocation name, ($to_i^{Long}$, $to_i^{Lat}$) are their respective long/lat values, and, $to_i^G$ signifying a geolocation token. This step plays a significant role in *N2T* because it identifies geolocation names, detects if the same geolocation name is visited multiple times, and preserves the listing order of the geolocation names as they appear in the text corpus.

*3.2.4 Geospatial Part of Speech (POS) Tagging*

It is an optional process that categorizes a token based on its semantics and position in a sentence. Applying NLP POS Tagging techniques to the tokens generated in the *Tokenization* step and using the Location Dimension dataset we create a list of 5-tuples (token, tag, geospatial flag, longitude, latitude).

*3.2.5 Trajectory Charting*

This step plots the trajectory using the *matplotlib-like* interface. The geospatial tokens that are generated by the previous two steps, are marked on a map using their longitude and latitude. A line is drawn connecting each location to the one that precedes it. The result is a trajectory that represents the travelled route over time as listed in the narrative.

In summary, *N2T* described in Fig. 1 is a suite of four methods:
1. *ST*: Significant Entity Tokenization
2. *ST+Geo_Aug: ST* + Geospatial Augmentation
3. *MWT*: Multi-Word Tokenization
4. *MWT+Geo_Aug: MWT*+Geospatial Augmentation

## 4 EXPERIMENTAL RESULTS

### 4.1 Description of the Dataset

Our text corpora is composed of several human trafficking narratives ($N_1 \ldots N_n$), a multi-word Lexicon Dimension, and a Location Dimension. The narratives are written by English speaking journalists, acquired from various news agencies and anti-trafficking organizations, and are published between 2015 and 2021. Their length varies between 605 and 4,212 words per narrative. There are 4 to 21 locations listed per narrative. Some location names consist of multiple words, while others contain non-ASCII characters. A survivor may have visited the same location multiple times. In addition, the trafficking narratives are location-ordered because they list location names according to the timeline of the events.

### 4.2 Ground Truth

To identify the ground truth from the text corpora we manually read each narrative, extracted the location names, and saved them sequentially in a ground truth structure $GT = (gt_i, | i = 1 \ldots n)$ where n = # of narratives in text corpora and $gt_i =$ ($geoloc_i^1 \ldots geoloc_i^m$) where $m$ # of geolocations in a trajectory.

### 4.3 Results

*4.3.1 Trafficking Narratives*

We executed each of the four methods of *N2T* using each narrative in the text corpus and generated a trajectory $Tr_i$ for each narrative $N_i$. Fig. 2 displays the trajectory $Tr_4$ generated by N2T using one migrant trafficking narrative. We compared every $Tr_i$ against the ground truth $gt_i$ and calculated the performance measures for each method applied. Our goal is to extract as many geospatial tokens as possible from text corpora. *ST* and *MWT* failed to extract most geospatial tokens; therefore we decided to enhance them by augmenting each of these two methods with Geospatial Dimension resulting in methods *ST+Geo_Aug* and *MWT+Geo_Aug* respectively. The latter methods leverage database techniques with NLP libraries, without the help of the geospatial tagging process. They reduce false positive outcomes to a value close to zero and increase the true positive values resulting in higher *F1-Score* than *ST* and *MWT*. *MWT+Geo_Aug* returns the highest

*F1-Score* compared with all other methods because it uses sentence tokenization, Multi-word Lexicon, and database techniques to recognize multi-word location tokens. This is illustrated in Fig. 3.

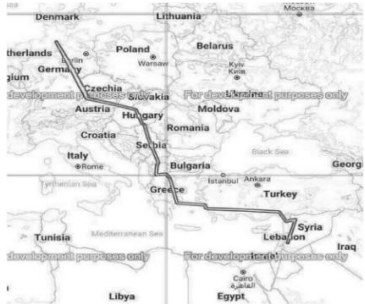
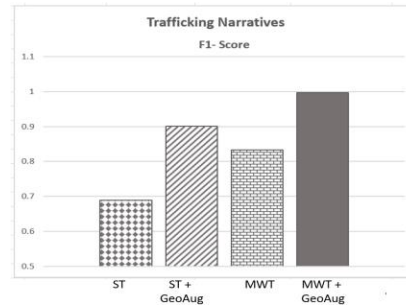

Fig 2. Plotted Trajectory of a trafficking narrative                   Fig. 3. F1-Score values for each method of *N2T*

*4.3.2 Comparing ILP to N2T*

We compared *MWT_Geo_Aug* to *ILP* system presented in [13] and found out that *MWT_Geo_Aug* outperforms *ILP* by 26.8% as shown in Table 1. Such higher performance is attributed to the thorough data preprocessing, the utilization of NLP libraries, and leveraging database techniques in *N2T MWT_GeoAug*, as opposed to minimal data preprocessing, the use of existing Readability Text Extractor, and relying on *Trie* data structure to extract city names in *ILP*. In addition, *ILP* is limited to cities with population > 15,000, whereas *N2T* renders all geographic names regardless of the population size.

Table 1: Performance Results of *ILP* vs. *N2T MWT+Geo_Aug*

| System | F1-Score |
| --- | --- |
| *ILP* | 0.785 |
| *N2T MWT+Geo_Aug* | 0.996 |

*4.3.3 Comparing BERT to N2T*

To address the homonymy of a given name referring to a location or not (e.g. Paris as a person, or a location), we trained a BERT model [17] on an NER annotated corpus containing 1.04 million words with plenty of geographic location names [18]. When tested it on the sentence "Mr. Paris drives a nice car", BERT identified Paris as person. However, when the title "Mr." was removed from the sentence, BERT identified Paris as a geographic location. *MWT_Geo_Aug* always identified Paris as a geographic location. We are currently enhancing *N2T MWT_Geo_Aug* to address the homonymy of names, whether a title is added before the name or not. In addition, *BERT* takes a significant amount of time to train a model, while *MWT_Geo_Aug* is considerably faster.

## 5 CONCLUSION AND FUTURE WORK

We tested each of the four methods of *N2T* using human trafficking narratives and identified that *MWT_Geo_Aug* outperforms all other methods. *N2T MWT_Geo_Aug* identifies more geographic location names than *ILP*. We addressed the homonymy concerning the POS tagging of words by comparing the performance of *MWT_Geo_Aug* to BERT's and found out that neither can differentiate if a given word is a location or not. The *N2T MWT_Geo_Aug* system is being enhanced to differentiate between homonyms. We are currently augmenting ontologies and post processing of the trajectory that we plan to present in the future.